# A Quantum Chemical Approach for the Characterization of the Interaction Potential of Propylene Oxide with Rare-Gas Atoms (He, Ne, Ar)

**Patricia R. P. Barreto** [1,*], **Ana Claudia P. S. Cruz** [1], **Henrique O. Euclides** [1], **Alessandra F. Albernaz** [2], **Federico Palazzetti** [3,*], **Fernando Pirani** [3]

[1] Instituto Nacional de Pesquisas Espaciais – São José dos Campos – São Paulo - Brazil.

[2] Universidade Nacional de Brasília – Instituto de Física - Brasília - Brazil.

[3] Università degli Studi di Perugia – Dipartimento di Chimica, Biologia e Biotecnologie, via Elce di Sotto, 8. Perugia, Italy

* Corresponding authors: prpbarreto@gmail.com; federico.palazzetti@unipg.it

**Abstract.** Propylene oxide is one of the simplest organic chiral molecules and has attracted considerable interest from the scientific community a few years ago, when it was discovered in the interstellar medium. Here, we report a preliminary study on the interaction between propylene oxide and rare-gas atoms, specifically He, Ne, and Ar. The interaction potentials as a function of the distance between the center-of-mass of propylene oxide and the rare-gas-atom are calculated for fourteen leading configurations at CCSD(T)/aug-cc-pVDZ level of theory. Symmetry Adapted Perturbation Theory has been employed for the analysis of the intermolecular potential, revealing that most of the contribution is given by dispersion and exchange forces.

# 1. Introduction

Propylene oxide is one of the simplest organic chiral molecules, making it an ideal candidate for studies concerning chiral selectivity mechanisms [1, 2]. It is characterized by a high vapor pressure and stability that makes this molecule suitable for scattering experiments, where high beam intensity and duty cycle are important. Its commercial availability is also remarkable for both the racemic mixture and the two enantiomers. It has been largely studied both experimentally, for example by synchrotron radiation [3] and molecular beams [4, 5], and from a computational point of view. For this latter aspect, see example the representation of the potential energy surfaces of propylene oxide with helium[6, 7] and the study of the structural isomers manifold of $C_3H_6O$ [8]. Recent discovery in the interstellar medium of this molecule [9] determined unusual interest, considering that numerous organic, even complex, molecules are continuously discovered [10–12].

In this work, we present the study by a quantum mechanical approach of the binary interaction between propylene oxide and the rare gas atoms He, Ne and Ar. *Ab initio* calculations have been carried out at different levels of theory to determine the most stable geometry of the molecule and the energy profiles of some significant configurations, called leading configurations (see also refs.[13, 14]), chosen upon physical and geometrical considerations. The *ab initio* points have been fitted with a fifth-order Rydberg potential function [15]. The Symmetry Adapted Perturbation Theory (SAPT) method has been employed to calculate electric properties of the systems propylene oxide – rare-gas atom, such as dipole moment, polarizability, quadrupole moment, ionization potential, electron affinity and proton affinity. These properties are employed for a preliminary analysis of electrostatic, exchange, induction and dispersion contributions to the intermolecular forces. The detailed study of prototype propylene oxide- noble gas systems is also basic to formulate the force fields (FF) and then to carry out molecular dynamics simulations in these and in other systems at increasing complexity and of major interest for the chiral recognition.

The article is structured as follows: in Section 2, we give a background of the quantum mechanical methods we employed and describe the leading configurations; in Section 3, we discuss the results; in Section 4, final remarks end the paper.

## 2. Background

### 2.1. Geometry optimization and calculation of potential energy profile of the leading configurations

The structure of propylene oxide has been optimized by Gaussian package [16] at various levels of theory (Table 1). One hundred energy points have been calculated at CCSD(T)/aug-cc-pVDZ by using MOLPRO [17] for a certain number of representative configurations, including the counterpoise correction to the basis set superposition error. As mentioned in Section 1, the choice of the representative configurations, namely leading configurations, relay on geometric and physical considerations of the system. The leading configurations of propylene oxide – He were discussed in Ref. [6] (see also [18]) and the same choice is applied to the systems propylene oxide – Ne and propylene oxide – Ar. Propylene oxide molecule is considered as built in a distorted tetrahedron (Figure 1), where the carbon atoms of $CH_3$ and $CH_2$ fragments, referred as C3 and C2, the oxygen, O, and the hydrogen, H, atom bound to the asymmetric carbon are the vertices. Fourteen configurations have been identified and indicated by the letters V, E and F, that stand for vertex, edge and face of the tetrahedron, respectively (Figure 2). In the four V configurations, the rare-gas-atom lies through the vertex – center-of-mass of propylene oxide direction. V1 is directed toward C3, V2 in correspondence of C2, V3 toward O, and V4 along the H – center-of-mass direction H. For the E configurations, the direction of the rare-gas-atom is referred to the line that connects the center-of-mass of the molecule and the center-of-mass of the edge of the tetrahedron, defined by the center-of-mass of two vertices. In the determination of the centers-of-mass of the edges involving C3 and C2, the whole masses of the groups $CH_3$ and $CH_2$ have been considered. Thus, E1 represents the configuration directed in correspondence of the center-of-mass of the edge C3-C2; E2 corresponds to the center-of-mass C3-O, E3 goes through C3-H, E4 through C2-O, E5 through C2-H, finally, and E6 is in correspondence of the center-of-mass of the edge O-H. The F configurations are defined as the directions that connect the rare-gas-atom to the centers-of-mass of the faces of the distorted tetrahedron and the center-of-mass of the molecule: in F1, the rare-gas is directed toward the center-of-mass of C3-C2-O, in F2 through C3-C2-H, in F3 through C3-O-H, and in F4 through the center of mass of C2-O-H.

The analytical form of the potential energy surface for the fourteen leading configurations is built by fitting a fifth degree generalized Rydberg function, V(R), to the calculated *ab initio* points:

$$V(R) = D_e \sum_{k=1}^{5} \left(1 + a_k(R - R_{eq})^k\right) \exp[-a_1(R - R_{eq})] + E_{ref} \qquad (1)$$

where $D_e$, $a_i$, $R_{eq}$ and $E_{ref}$ are adjustable parameters (see Supporting Information). The adjustable parameters have been obtained by a non-linear least-square procedure that minimized the differences between the analytic energies obtained by the potential function reported above and the *ab initio* energy points.

## 2.2. Symmetry adapted perturbation theory calculations

The SAPT method has been performed using the PSI4 code [19] at the CCSD(T)/aug-cc-pVDZ level of theory, on the configuration of minimum energy, among those considered in this work. In SAPT, the total

Hamiltonian for the molecule – atom system is partitioned as **H = F + V + W**, where **F = F$_A$ + F$_B$** is the sum of the Fock operators for monomers **A** and **B**, **V** is the intermolecular interaction operator, and **W = W$_A$ +W$_B$** is the sum of the Møller-Plesset operators. The Fock operators are treated as zero-order Hamiltonian, while the interaction energy is evaluated through a perturbative expansion of V given by:

$$E_{int} = \sum_{n=1}^{\infty}\sum_{j=1}^{\infty}\left(E_{pol}^{(nj)} + E_{exch}^{(nj)}\right) \qquad (2)$$

The polarization energies $E^{(nj)}_{pol}$ are identical to the corrections obtained in a regular Rayleigh-Schrödinger perturbation theory and the exchange corrections $E^{(nj)}_{exch}$ arise from the use of a global antisymmetrizer to force the correct permutation symmetry of the dimer wave function in each order, hence the name "symmetry adaptation". In this way, the interaction energy given by SAPT can be written as:

$$E_{SAPT} = E_{elst} + E_{exch} + E_{ind} + E_{disp} \qquad (3)$$

Using higher-order SAPT, as SAPT2+3, the electrostatic part is represented by:

$$E_{elst} = E_{elst}^{(10)} + E_{elst,resp}^{(12)} + E_{elst,resp}^{(13)} \qquad (4)$$

The superscript defines the order in V, while the subscript, *resp*, means that orbital relaxation effects are included. The exchange part is given by:

$$E_{exch} = E_{exch}^{(10)} + E_{exch}^{(11)} + E_{exch}^{(12)} \qquad (5)$$

The induction part is formulated as:

$$E_{ind} = E_{ind,r}^{(20)} + E_{ind,r}^{(30)} + {}^{t}E_{ind}^{(22)} + E_{exch-ind,r}^{(20)} + E_{exch-ind}^{(30)} + {}^{t}E_{exch-ind}^{(22)} - \delta_{HP}^{(2)} + \delta_{HH}^{(3)} \qquad (6)$$

The $\delta_{HF}^{(2)}$ and $\delta_{HF}^{(3)}$ terms take into account higher-order induction effects, the is the MP2 correlation part of not included in. Finally, the dispersion energy is given by:

$$E_{disp} = E_{disp}^{(20)} + E_{disp}^{(21)} + E_{disp}^{(30)} + E_{exch-dis}^{(20)} + E_{exch-dis}^{(30)} + E_{ind-disp}^{(30)} + E_{exch-ind-d}^{(30)} + \Delta CCD \qquad (7)$$

Where $\Delta CCD$ is an improved version of the CCD treatment of dispersion [20, 21].

## 3. Results and Discussion

In Table 1, we report the structural properties of the optimized geometry of propylene oxide, calculated at various levels of theory and compared with reference data. The CBS-QB3 method [22] presents overall the best agreement with reference data, for both distances and angles. Thus, the geometry optimized by CBS-QB3 is chosen for the calculation of the interaction potentials of the leading configurations as a function of the distance and for the analysis of the contributions to the intermolecular forces.

Potential energy profiles of the leading configurations of propylene oxide – He have been already presented and discussed in Ref. [6], where they have been compared with the Pirani potential function, aka Improved Lennard-Jones. Here, we use a fifth order Rydberg potential function, as discussed in Section 2.2 for the three systems. The single point energies of propylene oxide with the He, Ne and Ar are reported in Figures 3, 4 and 5, respectively. For propylene oxide – He, the minimum energy configuration is F1, at *ca.* 45 $cm^{-1}$ and 3.7 Å followed by E5, at a similar value of energy, but with an equilibrium distance of *ca.* 4.2 Å. The stability of F1 is probably due to hindrance steric reasons and to the polarizability effect of the groups $CH_3$, $CH_2$ and O atom. This is confirmed by the comparison with the systems propylene oxide – Ne and propylene oxide – Ar, where F1 is also in these cases the leading configuration with the minimum energy. Among the V configurations, V1 is the most stable for the three systems; V1, V2 and V3 present also similar values of the equilibrium, while V4, where the rare-gas atom is directed along the H, is less stable and at higher values of the equilibrium distance. Configurations E3 and E5, *i. e.* the "edges" $CH_3$ – H and $CH_2$ – H, also present higher values of the minimum energy and equilibrium distance, with respect to the other E configurations. Concerning F configurations, F2 and F3 present the highest repulsive character. It is important to point out that the potential energy profiles of the leading configurations have a similar trend for the three systems. As mentioned above, F1 is the configuration of minimum energy for the three systems. For propylene oxide – He, F1 presents a minimum energy of 36.9 $cm^{-1}$ at 3.6 Å; for propylene oxide – Ne, the depth of the potential well is *ca.* 60.7 $cm^{-1}$ at 3.7Å; finally, for propylene oxide – Ar, we have a minimum of 169.3 $cm^{-1}$ at 3.9Å.

In Table 2, we show the minimum energy and equilibrium distances for the three molecule – rare-gas-atom systems calculated at the CCSD(T)/aug-cc-pVDZ level of theory in the F1 configuration, comparing the results with and without counterpoise correction for the basis set superposition error. The result shows that the counterpoise correction tends to increase the equilibrium distance of about 0.2 Å for all the systems. However, the most significant effect is found in the minimum energy, where the counterpoise correction considerably decreases the attractive component of the potential. For the propylene oxide – He systems, the potential calculated without counterpoise correction is twice as attractive than that calculated with the counterpoise correction, while for propylene oxide – Ne and – Ar systems this difference is even higher. We must note that the lower values of energy are probably the result of an overcorrection effect, especially considering the limitation of the basis set. Table 3 reports the difference between the minimum energy and equilibrium distance calculated *ab initio* with and without counterpoise correction, confirming the trend already observed for the F1 configuration.

Table 4 reports the electric properties: dipole moment, polarizability, quadrupole moment, ionization potential, electron affinity and proton affinity, for the three systems investigated. For propylene oxide the zero-point energy is also presented. These properties have been then employed to analyze the contributions to the interaction potentials [23–26].

Figure 5 compares the energy profiles at different SAPT correction levels for the F1 configuration. Except for the Hartree-Fock (HF), that presents repulsive character, all the corrections present similar energy profiles. For propylene oxide – He, the minimum has been obtained at SAPT2+(CCD) level, for propylene oxide – Ne at SAPT2+, and for propylene oxide - Ar at SAPT0.

Figure 6 shows the interaction potential and the related contributions of the three systems for the F1 configuration calculated at SAPT2+(3)(CCD) level of theory. As expected, the equilibrium distance and the minimum energy increases with the atomic mass of the rare gas atom. The plots show that a fundamental role is played by dispersion forces, whose attractive character is already present in the long-range distance and increases by passing from He, to Ne and Ar. The exchange contribution, which presents a purely repulsive character, also depends on the atomic mass of the rare gas atom and its effect, as well as that of dispersion forces, is present at increasing distances moving from He to Ar. Induction and electrostatic contributions have been also considered in this treatment; their contribution is negligible in the long- and medium- range of distances. For what concerns the electrostatic contribution, it becomes attractive in the vicinity of the potential-well, where the dispersion and exchange forces play the most important role, thus the contribution of the electrostatic interactions is almost negligible. Even lower is the effect of induction forces, whose attractive character becomes relevant in the short range of distances, where the repulsive contribution of the exchange forces conceals all the other contributions. Table 4 compares the minimum obtained from *ab initio* and SAPT calculation for the configuration F1. The energies obtained by the SAPT and *ab initio* method with counterpoise correction present a good agreement, with an error lower than 1% in energy

## 4. Final Remarks

The interaction potential of fourteen configurations of the systems propylene oxide – rare-gas-atoms (He, Ne and Ar) have been calculated at the CCSDT(t) level of theory, using the aug-cc-pVDZ basis set. The geometry of propylene oxide, modeled as rigid, has been optimized at various levels of theory and CBS-Q3 gave the best agreement with reference data. The analytical form of the potential energy surface is generated by fitting the single point energies with a fifth-order generalize Rydberg potential function. The potential energy profiles of the three systems present a similar trend and F1 is the most stable leading configuration for all the three rare-gases. As expected, the minimum energy of the leading configuration become more negative passing from the He to the Ne ad the Ar. Similarly, the equilibrium distances become larger.

The SAPT method has been applied for a preliminary analysis of the contributions to the interaction potentials from the calculated electric properties. The three systems present a similar behavior; the repulsive contribution is given exclusively the exchange interaction, while the dispersion forces play the most important role for what concerns the attractive character in the portion of distances we have considered.

Future developments, addressed to build up FFs useful for molecular dynamics simulations of chiral effects, will involve experimental investigations on propylene oxide – rare gas collision processes, useful for an accurate analysis of the nature and for a general formulation of the interaction potentials (see for example [27–29]).. For a further validity test of predicted intermolecular interactions it is also of fundamental importance their comparison with results of quantum mechanical calculations that make use of more accurate basis set functions.

**Acknowledgments.** Federico Palazzetti acknowledges the Italian Ministry for Education, University and Research, MIUR, for financial supporting through SIR 2014 "Scientific Independence for young Researchers" (RBSI14U3VF).

**Figure 1**

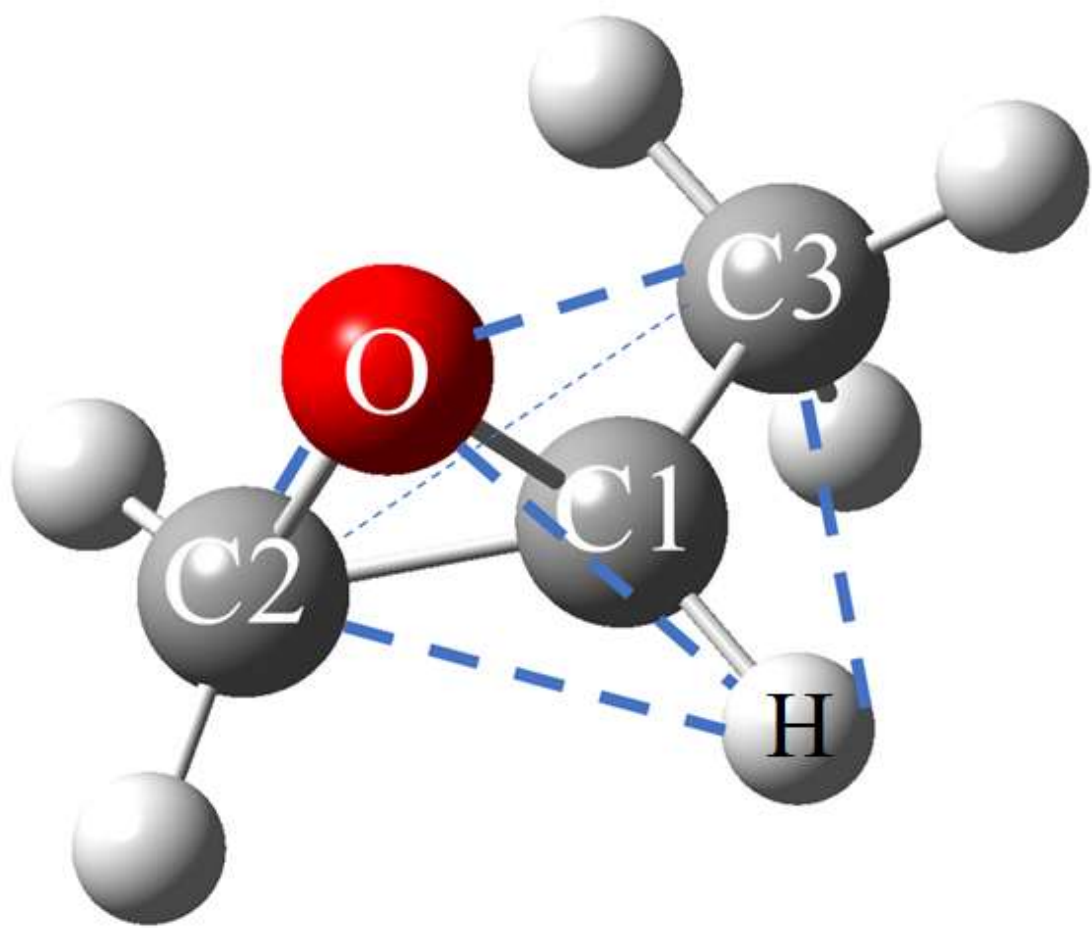

**Figure 2**

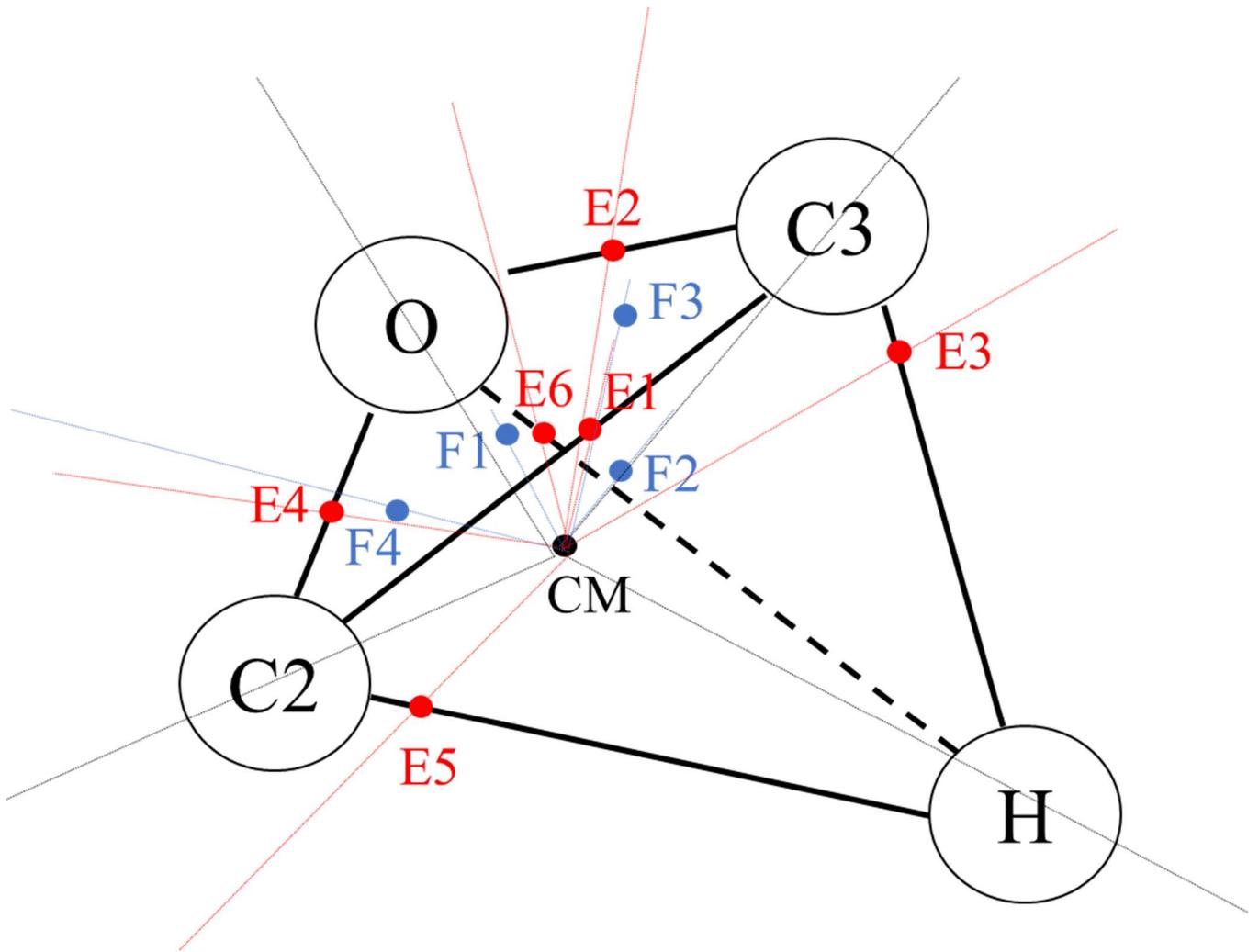

**Figure 3**

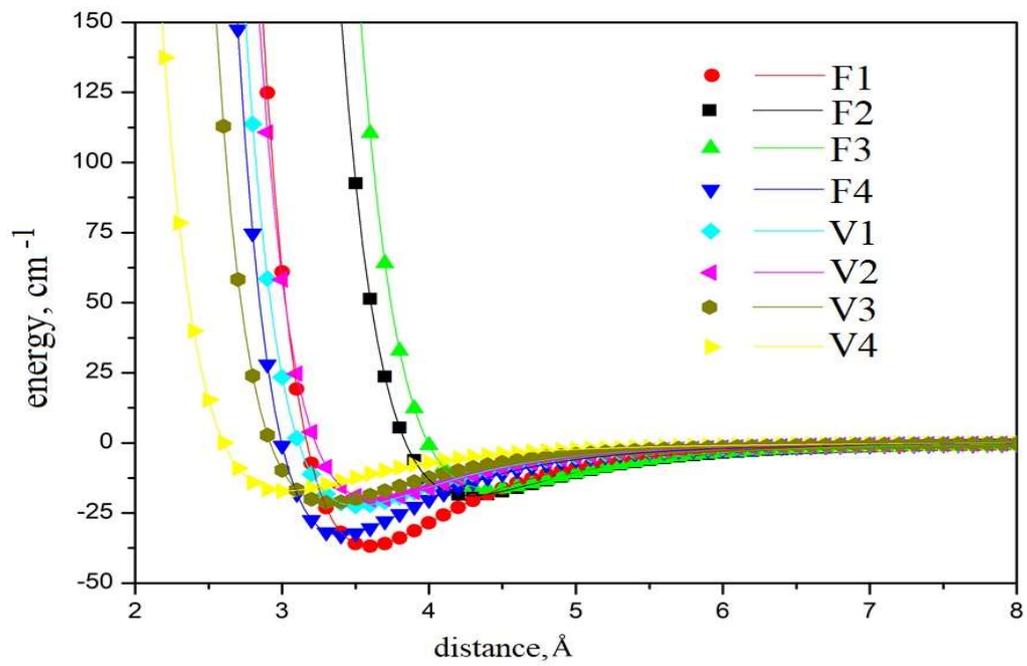

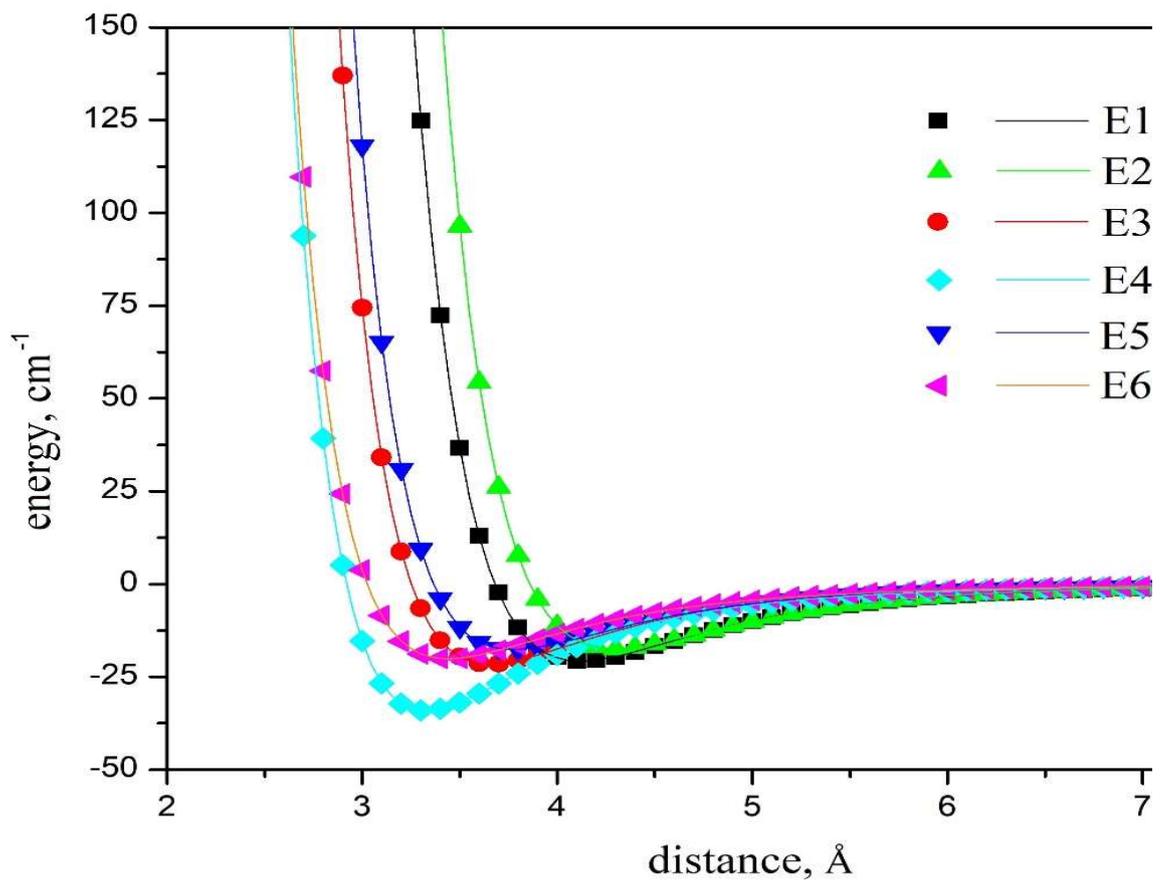

**Figure 4**

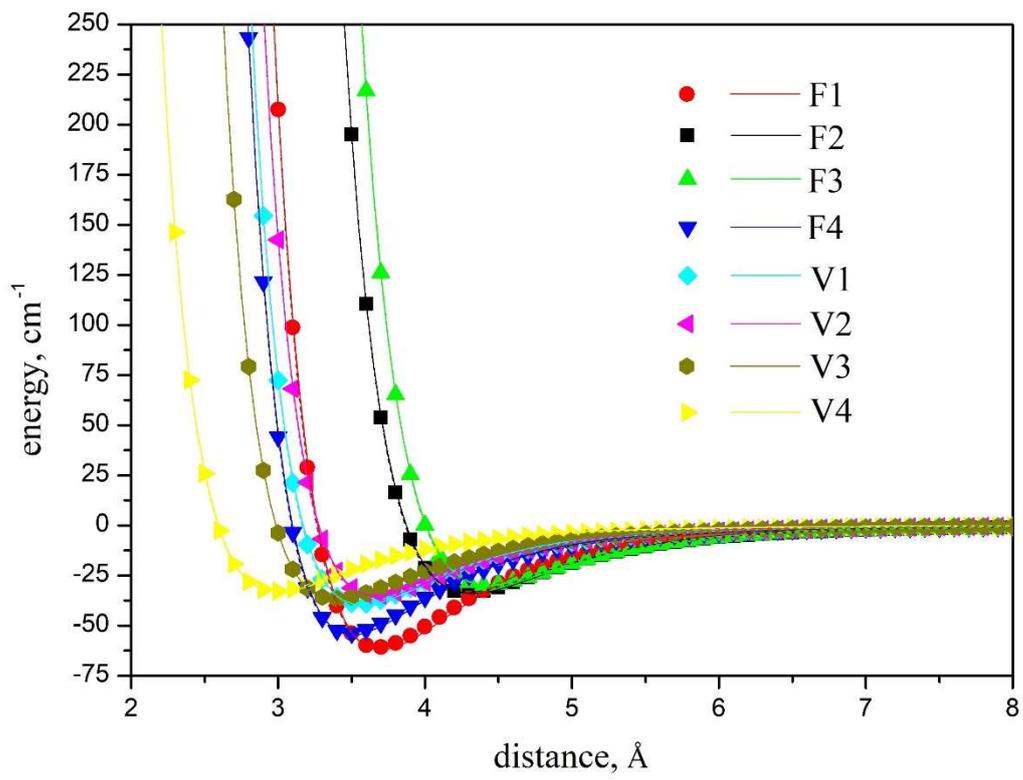

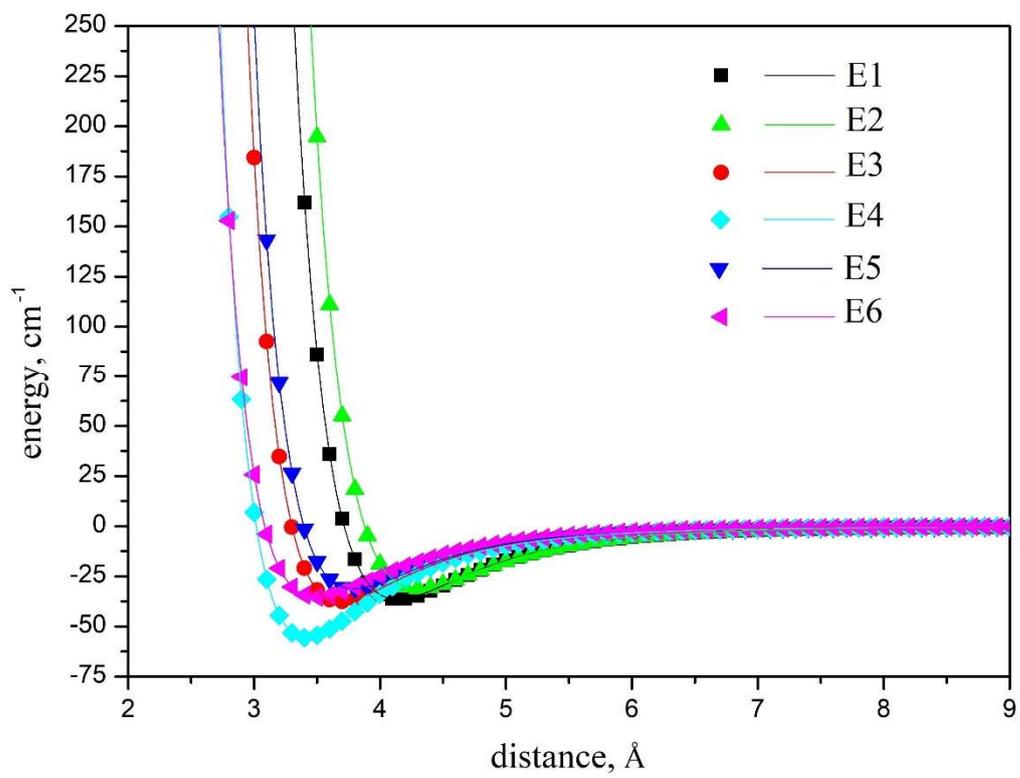

**Figure 5**

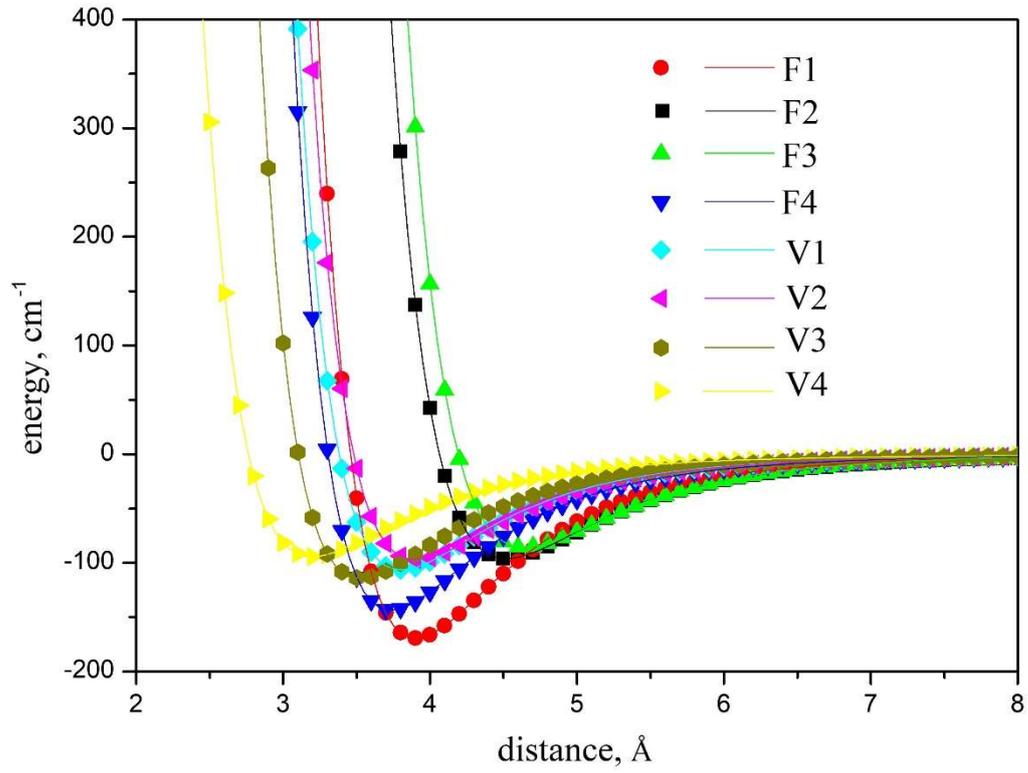

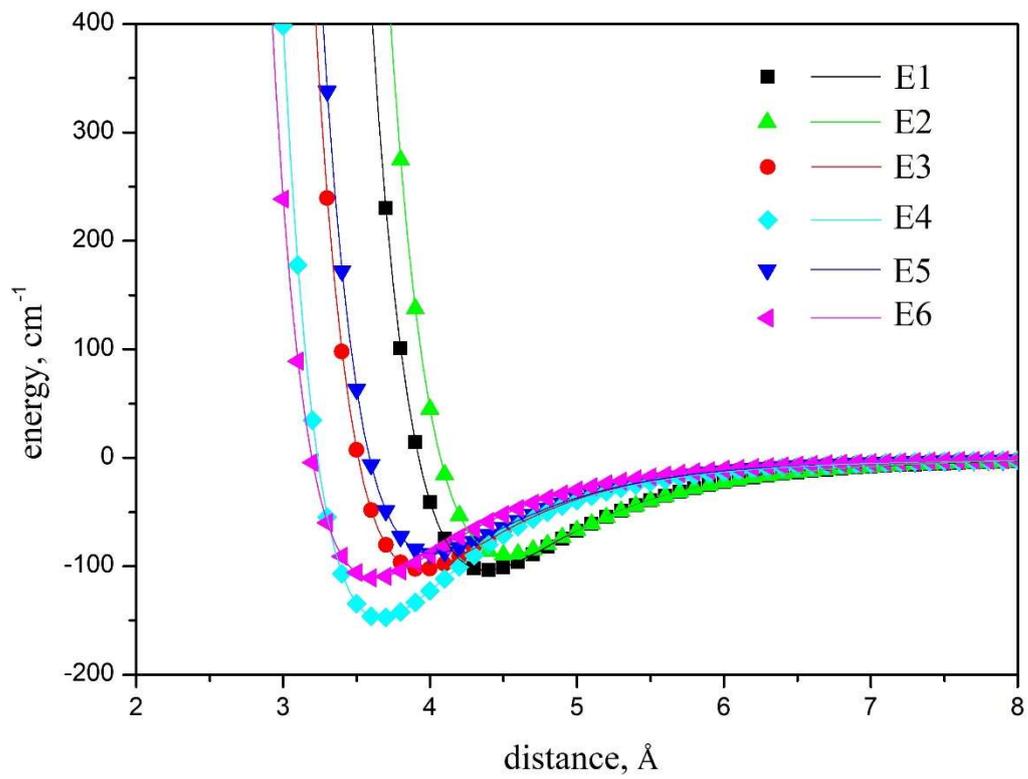

**Figure 6**

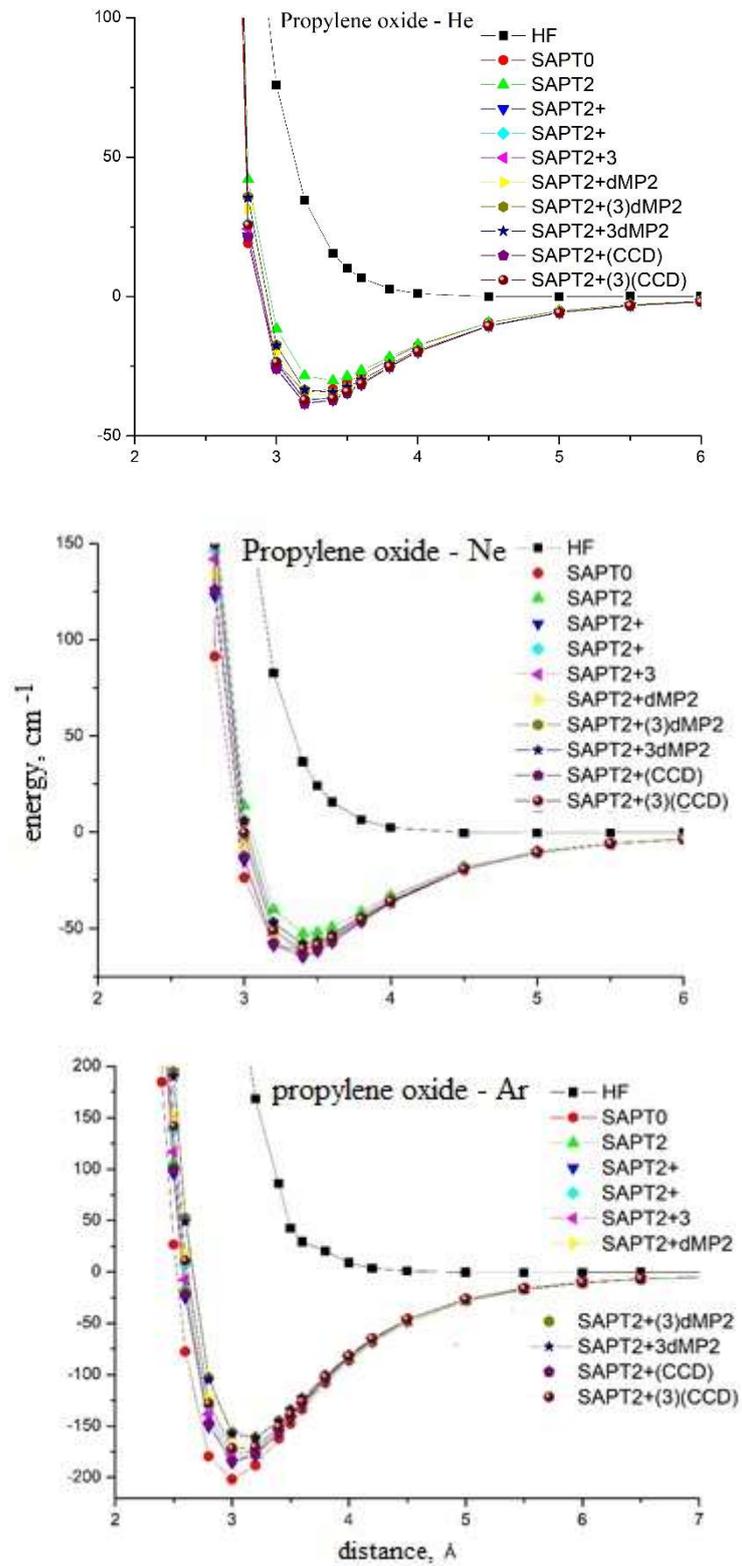

**Figure 7**

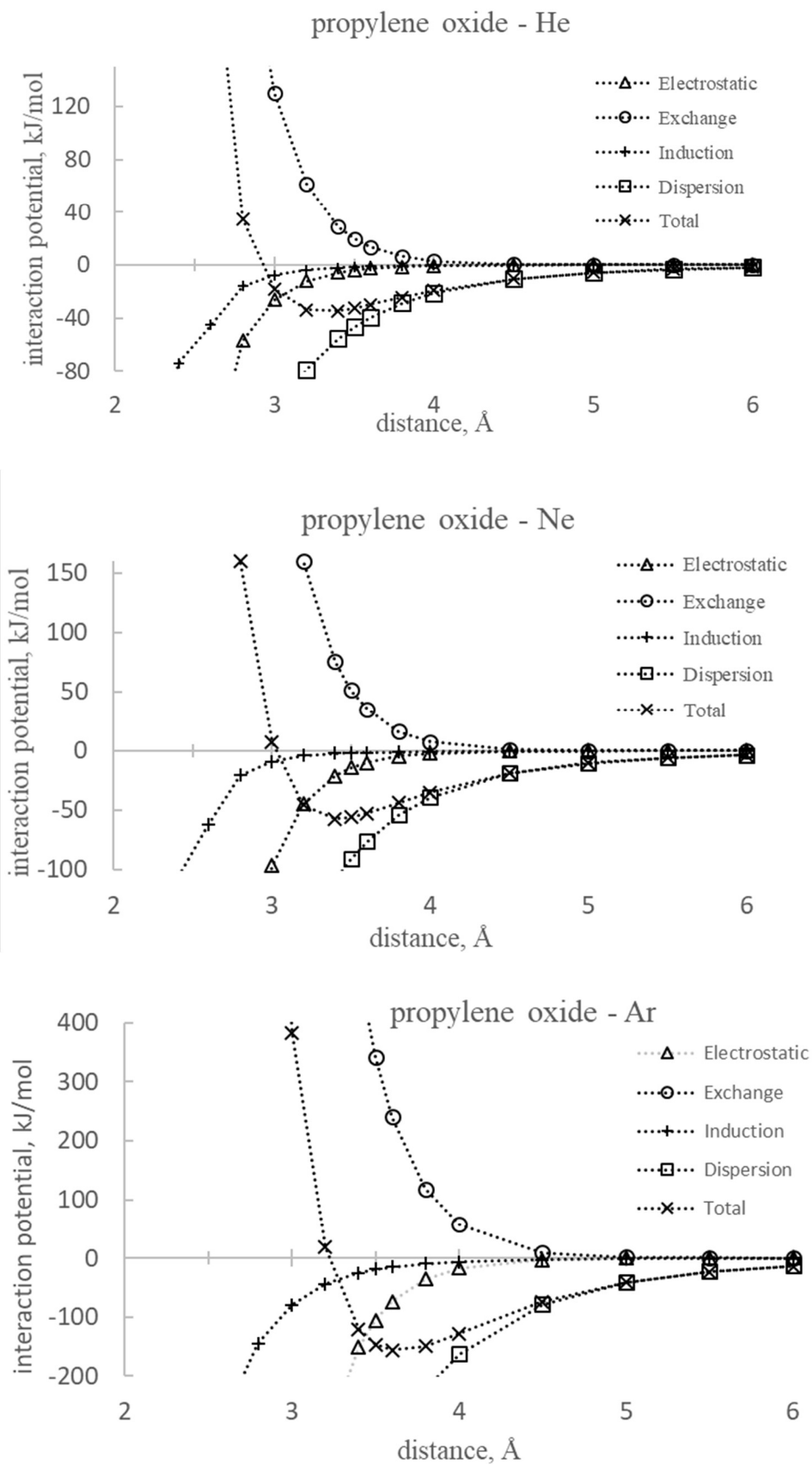

# Captions

**Figure 1.** The propylene oxide molecule: in white the hydrogen atoms; in grey the carbons C1, C2 and C3; in red the oxygen.

**Figure 2.** The propylene oxide is represented as a distorted tetrahedron, whose vertices are the H, O, C2, and C3; the edges are indicated by thick black lines. The leading configurations are classified as V, E and F. The four V configurations correspond to the rare-gas atom directed through the vertices of the tetrahedron and the center-of-mass of the molecule (black dot); in V1 the rare-gas atom is directed through C3, in V2 through C2, in V3 through O, and in V4 through H (the direction of the V configurations are indicated by a black line). The six leading configurations E correspond to the direction (red line) connecting the center-of-mass of the molecule and the centers-of-mass of the six edges of the tetrahedron (red dots). The four leading configurations F correspond to the direction (blue line) connecting the center-of-mass of the molecule and the centers-of-mass of the faces of the tetrahedron (blue dots).

**Figure 3.** Potential energy curves of propylene oxide – He of the fourteen leading configurations. The *ab initio* points are indicated by symbols and the Rydberg fitting by lines.

**Figure 4.** Potential energy curves of propylene oxide – Ne of the fourteen leading configurations. The *ab initio* points are indicated by symbols and the Rydberg fitting by lines.

**Figure 5.** Potential energy curves of propylene oxide – Ar of the fourteen leading configurations. The *ab initio* points are indicated by symbols and the Rydberg fitting by lines.

**Figure 6.** Interaction potential as a function of the distance, at different SAPT corrections, up to SAPT2+(3)(CCD), for the F1 configuration. Propylene oxide - Helium (upper panel), Neon (medium panel) and Argon (lower panel).

**Figure 7.** Contributions to the interaction potential determined at SAPT2+(3)(CCD) level for the F3 configuration: Helium (upper panel), Neon (middle panel), Argon (lower panel).